\documentclass{emulateapj}
\usepackage{graphicx}

\newcommand{\rms}{{\em rms}}
\newcommand{\twosig}{$2\sigma$}
\newcommand{\lstar}{$L^*$}

\newcommand{\mstar}{$M^*$}

\begin{document}

\title{ Star formation in the early universe: beyond the tip of the iceberg}

\author{N.~R.~Tanvir\altaffilmark{1},
A.~J.~Levan\altaffilmark{2}, A.~S.~Fruchter\altaffilmark{3}, J.~P.~U.~Fynbo\altaffilmark{4}, J.~Hjorth\altaffilmark{4}, 
K.~Wiersema\altaffilmark{1},  M.~N.~Bremer\altaffilmark{5}, J.~Rhoads\altaffilmark{6}, P.~Jakobsson\altaffilmark{7}, 
P.~T.~O'Brien\altaffilmark{1},
E.~R.~Stanway\altaffilmark{2},
D.~Bersier\altaffilmark{8},
P.~Natarajan\altaffilmark{9},
J.~Greiner\altaffilmark{10},
D.~Watson\altaffilmark{4},
A.~J.~Castro-Tirado\altaffilmark{11},
R.~A.~M.~J.~Wijers\altaffilmark{12},
R.~L.~C.~Starling\altaffilmark{1},
K.~Misra\altaffilmark{3},
J.~F.~Graham\altaffilmark{3},
C.~Kouveliotou\altaffilmark{13}
}
\altaffiltext{1}{Department of Physics and Astronomy, University of Leicester, University Road,
Leicester, LE1 7RH, UK}
\altaffiltext{2}{Department of Physics, University of Warwick, Coventry, CV4 7AL, UK}
\altaffiltext{3}{Space Telescope Science Institute, 3700 San Martin Drive, Baltimore, MD 21218, USA}
\altaffiltext{4}{Dark Cosmology Centre, Niels Bohr Institute, University of Copenhagen, Juliane Maries Vej 30, DK-2100 Copenhagen \O, Denmark}
\altaffiltext{5}{HH Wills Physics Laboratory, University of Bristol, Tyndall Avenue, 
Bristol, BS8 1TL, UK}
\altaffiltext{6}{School of Earth and Space Exploration, Arizona State University, USA}
\altaffiltext{7}{Centre for Astrophysics and Cosmology, Science Institute, University of Iceland, Dunhagi 5, 107 Reykjav\'ik, Iceland}
\altaffiltext{8}{Astrophysics Research Institute, Liverpool John Moores University, Liverpool, UK}
\altaffiltext{9}{Department of Astronomy, Yale University, New Haven, CT 06511-208101, USA; Department of Physics, Yale University, New Haven, CT 06520-208120, USA}
\altaffiltext{10}{Max-Planck-Institut f\"ur extraterrestrische Physik, Garching bei M\"unchen, Germany}
\altaffiltext{11}{Instituto de Astrof'sica de Andaluc\'ia (IAA-CSIC), Glorieta de la Astronom\'ia s/n, 18.008 Granada, Spain}
\altaffiltext{12}{Astronomical Institute 'Anton Pannekoek', PO Box 94248, 1090 SJ Amsterdam, the Netherlands}
\altaffiltext{13}{Space Science Office, VP62, NASA/Marshall Space Flight Center, Huntsville, AL 35812, USA}
\email{nrt3@star.le.ac.uk}


\begin{abstract}
We present late-time {\em Hubble Space Telescope (HST)} imaging of 
the fields of six {\em Swift} gamma-ray bursts (GRBs) lying at
$5.0 \lesssim z \lesssim 9.5$.
Our data include
very deep observations  of the field of 
the most distant spectroscopically confirmed burst, GRB\,090423, at $z=8.2$.
Using the precise positions afforded by 
their afterglows we can place 
stringent limits on the luminosities of  their host galaxies.
In one case, that of GRB\,060522 at $z=5.11$, 
there is a marginal excess of flux close to the GRB position which may be
a detection of a host at a magnitude $J_{\rm AB}\approx28.5$.
None of the others are significantly detected
meaning that all the hosts lie 
below \lstar\ at their respective redshifts, with star formation rates SFR~$\lesssim 4$ M$_{\odot}$\,yr$^{-1}$ in all cases. 
Indeed, stacking the five fields with WFC3-IR data we conclude a mean SFR~$<0.17$ M$_{\odot}$ yr$^{-1}$ per galaxy.
These results support the proposition that the bulk of star formation, and hence integrated UV luminosity, at high redshifts arises in 
galaxies below the detection limits of  deep-field observations. 
Making the reasonable assumption that GRB rate is proportional to UV luminosity at early times
allows us to compare our limits with expectations based on galaxy luminosity functions derived
from the Hubble Ultra-Deep Field (HUDF) and other deep fields.
We infer that a luminosity function which is evolving rapidly towards steeper faint-end slope ($\alpha$) and
decreasing characteristic luminosity (\lstar), as suggested by some other studies, is consistent with our observations,
whereas a non-evolving LF shape 
is 
ruled out at 
$\gtrsim90$\% confidence.
Although it is not yet possible to make stronger statements,
in the future, with larger samples and a fuller understanding
of the conditions required for GRB production,  
studies like this 
hold great potential
for probing the nature of star formation,
the shape of the galaxy luminosity function, and the supply of ionizing photons in the early universe.

\end{abstract}

\section{Introduction}
\label{intro}

The drive to locate and characterize the earliest gravitationally 
bound objects in the Universe -- the first proto-galaxies,
population III \& II stars and the black holes they produced
-- is a central theme of contemporary cosmology.
It is intimately tied to the quest to understand how the Universe at large was reionized, 
quite possibly by the ultra-violet (UV) flux produced by these first, and still enigmatic objects.  
Major  investments of time  on large facilities 
continue to be spent on these ambitious goals, but progress has 
been hard-won. 
Historically, quasars were the key high-redshift beacons, thanks to their great  luminosities,  but the space density of bright quasars drops
above $z\sim4$ and 
to date the most distant quasar is located at $z = 7.1$
\citep{Mortlock11}.
Galaxies are, of course, much more common than quasars, and high-$z$ galaxies are likely to be present in moderate numbers
in very deep near-infrared (nIR) observations. Indeed, recent nIR observations in the Hubble Ultradeep
Field (HUDF) have unveiled $>100$ 
candidate $z>7$ galaxies \citep[e.g.,][]{Bouwens11a,bunker10,mclure10}
discovered via the Lyman-dropout technique. 
The challenge  of
studying these sources lies in their extreme faintness (often $>$28th magnitude), which 
generally  means only photometrically-derived redshift estimates are possible.
Furthermore, distinguishing young, high-$z$ galaxies from old or dusty populations at moderate redshifts, 
or even Galactic brown dwarfs, 
becomes
increasingly difficult 
when approaching the sensitivity limit of the data.
Although spectroscopic redshifts have been 
determined for a small number of $z>7$ galaxies \citep{vanzella11,ono11}, this is only
possible if the galaxy is a strong Ly$\alpha$ emitter (and the emission is not
completely absorbed by a partially neutral
intergalactic medium), and requires a major investment of time for each target.
Even in the era of extremely large telescopes,
many such sources may be too faint for direct spectroscopic redshift confirmation.

The extreme luminosity of gamma-ray bursts (GRBs) makes them potentially powerful probes of the early Universe, 
a utility which has been widely touted since the discovery of the first
afterglows \citep[e.g.,][]{wijers98,lamb00,Tanvir:2007}. Current technology has the capability to detect the prompt, and afterglow emission 
for bright
GRBs out to $z\sim20$, should GRBs exist at this epoch \citep{gou04,racusin08,bloom09}, and they have several advantages over
alternative methods for the detection and study of high$-z$ objects. 
In the first instance, 
the intrinsically smooth power-law spectra of the afterglows makes them ideal backlights 
for absorption diagnostics: not only providing redshifts, but in principle chemical enrichment, hydrogen column
densities \citep[e.g.,][]{vreeswijk04,prochaska07,Fynbo:2009lr}, extinction and dust laws in the hosts \citep[e.g.,][]{zafar11b,schady12}, 
and even probing the state of the intergalactic medium (IGM) \citep[e.g,][]{miralda98,mcquinn08}.
Secondly, they pinpoint the positions of their hosts, and once the afterglow has faded we can search for the host galaxy with a redshift in hand
from spectroscopic (or photometric) observations of the afterglow. 
Finally, long-duration GRBs are produced during core-collapse events \citep[e.g.,][]{hjorth03,stanek03,pian06}, 
and so trace the locations of at least some component of massive star formation.
Importantly, GRBs allow
absorption redshift measurements for galaxies at times too faint to
be seen in deep imaging even with the {\em Hubble Space Telescope} ({\em HST}\,) \citep[e.g.,][]{berger02,hjorth03b,thoene10}.

The early populations of massive stars and the proto-galaxies in which they reside are
thought to be major producers of ionizing photons, and also give rise to GRBs.  
Hence, mapping GRB space density as a function
of cosmic time should trace
this early star formation and so provide a key to assessing its
contribution to the reionization of the IGM at $z>6$.
Since GRB progenitors are individual stellar systems, their
hosts should sample  the whole (star-forming) galaxy luminosity function (LF), rather than just the bright end, avoiding
the limitation inherent in flux limited samples \citep[e.g.,][]{Jakobsson05}. 
This is particularly important at very high-redshift, when galaxies were small and faint, and only the tip of the 
LF can be probed directly.

However, GRB-based studies are subject to
their own difficulties.
As with any sources beyond $z \sim 6$,
in addition to the increasing luminosity distance, 
we have to contend with the difficulties of working in the  nIR.
In addition, because GRB
afterglows fade there is only a narrow window of opportunity in which
observations can be pursued, so signal-to-noise cannot be built up over many
nights of observation. Hence, a major constraint in their detection and study 
at high-$z$ is the availability of large aperture telescopes which can respond
rapidly with appropriate nIR instrumentation.
It is also clear that  high-redshift GRBs detectable to the limits of the {\it Swift}
satellite are rather rare, and therefore many bursts have to be observed
to find the few at $z>5$ \citep[e.g.,][]{Jakobsson12}.
Finally, studies of GRBs at lower redshift have indicated that they are generally found in 
small-to-moderate size, low-to-moderate metallicity galaxies with high specific star-formation rates   
\citep[$= SFR/M_{\rm gal}$; e.g.,][]{christensen04,castroceron06,modjaz08,chen09,savaglio09,svensson10},
and rarely in intensively star-forming far-IR-bright galaxies \citep{tanvir04,lefloch06,michalowski08}.
Coupled with theoretical arguments \citep[e.g.,][]{yoon05,woosley06} and spectroscopic studies of 
their environments \citep[e.g.,][]{Fynbo:2009lr,levesque10a}, this has led to the suggestion that GRBs may 
be more common and/or brighter at relatively low metallicities \citep[e.g.,][]{wolf07}.
Such a dependence could potentially help explain the apparent increase in GRB-rate to SFR 
ratio between $z\sim0$ and $z\sim5$ seen in various studies \citep[e.g.,][]{natarajan05,yuksel08,robertson12}.
However, the picture is not a simple one: there is a selection bias against finding heavily 
extinguished afterglows \citep[e.g.,][]{perley09,kruehler11}, and at least some GRBs appear to be 
produced in very dusty, massive star-bursting galaxies 
\citep[e.g.,][]{chen10,kupcuyoldas10,hashimoto10,svensson12}, whilst others have been found 
in solar or even super-solar metallicity environments \citep{levesque10b,savaglio12}.
Furthermore, comparison of GRB luminosity and host metallicity at $z<1$ reveals no obvious 
correlation \citep{levesque10c}.
In any event, it is likely that galaxies with  properties similar to 
typical GRB hosts predominate at early times,  and indeed the small number of higher-redshift
afterglows for which accurate metallicities have been derived show a range that matches
well predictions from galaxy evolution simulations that include no bias in GRB production \citep{pontzen10}.
Thus, we expect GRBs to be
a good tracer of the bulk of high 
redshift star formation \citep{fynbo06,fynbo08,kocevski09}, and we proceed under that 
assumption in this paper.

Several very high-$z$ bursts have been identified by {\it Swift} to date, and have begun to realise their potential as probes
of the early universe.
The afterglow of GRB\,050904
at $z=6.3$ was brighter than magnitude $J=17.5$ even a few hours after the burst \citep{haislip06}, and spectroscopy would have been
routine with a 4--8\,m class telescope. Indeed, ultimately the redshift measurement came from a Subaru 
spectrum taken some 3 days after the event \citep{kawai06}. Even the intrinsically rather faint afterglows of GRB\,080913 
\citep[$i=20$; ][]{greiner08b}
or  GRB\,090423 
\citep[$J=21$; ][]{tfl09}
allowed for spectroscopic redshifts with 8\,m instrumentation, despite the data being acquired more than 12 hours after the burst in the latter case.

Until now, the most distant GRB host detected (with {\em Spitzer}) was
that of GRB\,060510B at $z=4.94$ \citep{chary07}.
In this paper we present deep, late-time {\em Hubble
Space Telescope (HST)} observations 
of the fields of six of the 
most distant gamma-ray bursts, all at $z>5$.
Only one host galaxy, that of GRB\,060522, could be marginally detected,
and the deep limits for the others
imply that all lie close to or below the characteristic luminosity, \lstar\, at their respective redshifts. 
As we show in this paper, the non-detections of the
hosts place important constraints on their total luminosities and star formation rates.
Since long-duration GRBs trace (at least) some component of 
star formation, a survey of even a small number of 
GRBs at high redshifts provides a census of
the locations and galactic environments of star formation 
at early times.
This means that GRBs have the potential ultimately to constrain the faint end  of the galaxy LF 
at $z > 7$, which is
crucial for understanding the reionization 
of the Universe, thought to occur predominantly at $8 < z < 12$ \citep{komatsu:2011tg}. 
UV photons produced by
massive stars are 
widely considered the most likely
driving force for reionization \citep{loeb09,faucher09}, but even ultra-deep surveys cannot
currently 
quantify the (likely dominant) contribution 
of intrinsically faint galaxies.

Throughout this paper we use the AB-magnitude system, and adopt a $\Lambda$-CDM cosmology
with $H_0=72$\,km\,s$^{-1}$\,Mpc$^{-1}$ and $\Omega_M = 0.27$, $\Omega_{\Lambda} = 0.73$.

\begin{deluxetable*}{llcccc}
\footnotesize
\tablecolumns{4}
\tablewidth{0pt}
\tablecaption{Summary of burst sample and astrometric ties}
\tablehead{\colhead{GRB} & \colhead{$z$} & \colhead{$z$-ref} &
\colhead{Obs. summary}&  \colhead{Astrometry} & \colhead{Accuracy (\arcsec)}}
\startdata
060522 & 5.11 & 1& WFC3:F110W & TNG & 0.06 \\
060927 & 5.47 & 2 & NIC3:F160W, WFC3:F110W & VLT/FORS2 & 0.07\\
050904 & 6.29 & 3& ACS:F850LP & Gemini-S/GMOS & 0.06 \\
080913 & 6.73 & 4& WFC3:F160W & VLT/FORS2 & 0.08 \\
090423 & 8.23 & 5 & WFC3:F125W,F160W & VLT/HAWK-I & 0.03 \\
090429B & 9.4* & 6 & WFC3:F105W,F160W & Gemini/NIRI & 0.06 \\
\hline
\enddata
\tablecomments{
References (1)  \citet{cenko06}, (2) \citet{ruizvelasco07}, 
(3) \citet{kawai06}, 
(4) \citet{patel10}, (5) \citet{tfl09},
(6) \citet{Cucchiara11}. *Photometric redshift estimate.}
\label{astro}
\end{deluxetable*}

\section{GRBs in the sample}
\begin{figure*}[ht]
\begin{center}
\resizebox{17truecm}{!}{\includegraphics[angle=90]{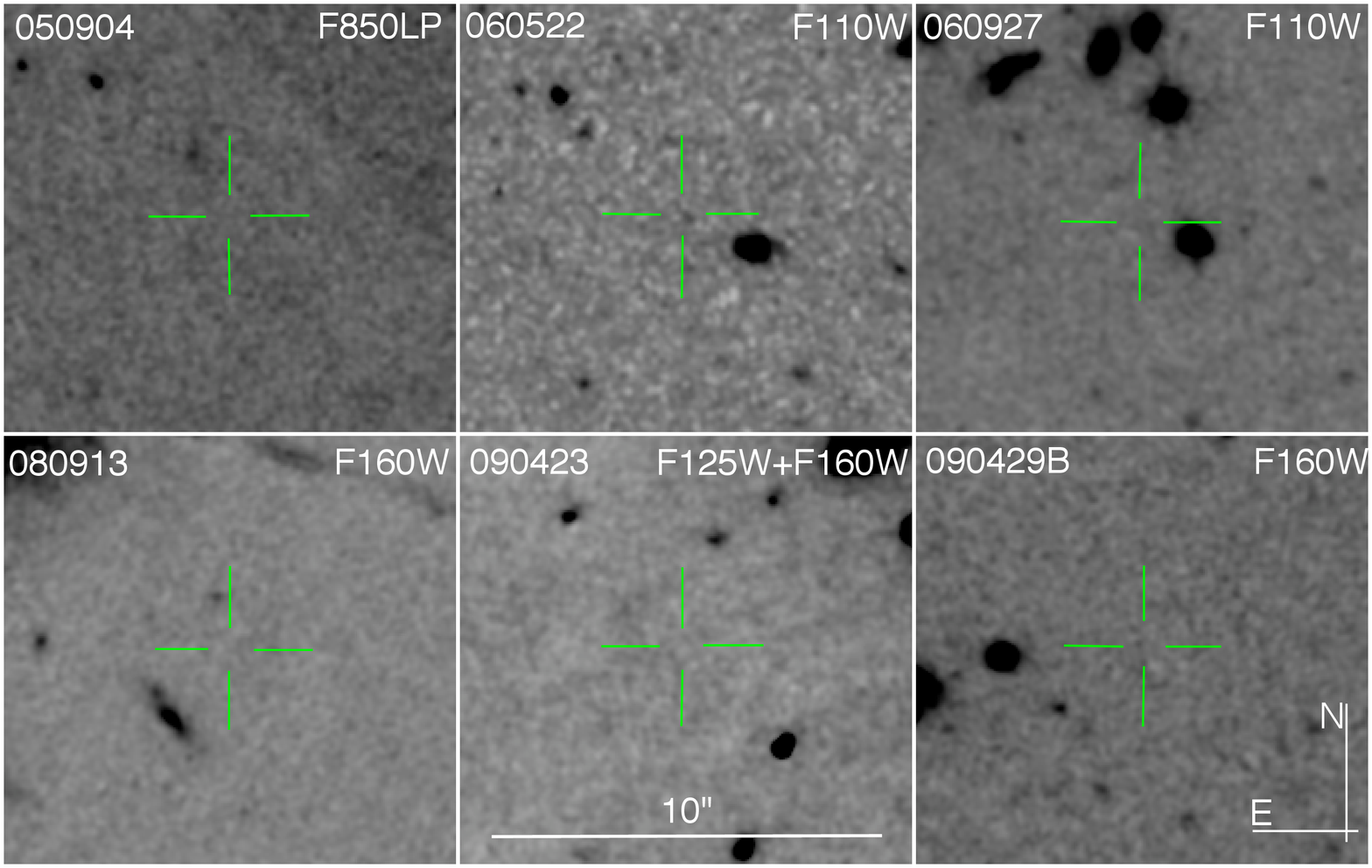}}
\end{center}
\caption{A mosaic of the host galaxy fields of the $z>5$ GRBs  in our study. The filters of the observations are indicated in each panel, while 
the cross-hairs represent the location of each burst as determined from astrometry with ground based images containing the afterglow (see Table~\ref{astro}
for more details).  The images have been lightly smoothed to allow the eye to see fainter features.}
\label{fig1} 
\end{figure*}

Our sample comprises six of the most distant bursts detected by {\em Swift}, including
all five bursts detected to date that have firm spectroscopic redshifts above $z=5$. 
They are
GRBs\,050904 
\citep[$z=6.29$; ][]{kawai06}, 
060522 \citep[$z=5.11$; ][]{cenko06},
060927 \citep[$z=5.47$; ][]{ruizvelasco07}, 
080913 \citep[$z=6.73$; ][]{greiner08b,patel10},
 090423  \citep[$z=8.23$; ][]{tfl09,sdc09}
 and GRB\,090429B 
\citep[$z_{\rm phot} = 9.4$; ][]{Cucchiara11}. 
Here, we briefly summarise details of each GRB, and the available  {\em HST}
observations of each field, which were mostly obtained after the afterglow
should have faded beyond detectability (the possible exception
is GRB\,050904 as discussed below).
The majority of our data comes from the new Wide Field Camera 3 Infrared channel (WFC3 IR). In all of these cases we used the standard
flat-fielded observations from the {\it HST} archive\footnote{archive.stsci.edu}, and corrected these for geometric distortion, creating a 
stack of individual images using {\tt multidrizzle}. The final pixel scale for these well dithered observations is set as 0\farcs05 per pixel,
with the {\tt pixfrac} set to 0.7.  For observations with ACS we  drizzled to the same final pixel scale, which broadly retains the native 
pixel scale of the instrument.  For GRB~060927, we also present NICMOS observations of the host, as detailed below. 
Cut-outs of the images around the locations of the GRB positions are shown in Fig.~\ref{fig1}.

The exact locations of each burst on the {\em HST} images were found relative to field sources using
ground-based
images of their afterglows.
Details of the data obtained for each burst and the {\it rms} accuracy of the
astrometric calibration are summarised in Table~\ref{astro}.

For photometry, following \citet{Bouwens:2010ly}, we employed  0\farcs4 diameter apertures centred at the
locations of the afterglows (for reference, 0\farcs4
corresponds to a physical scale of $\approx2.1$\,kpc at $z=7$). 
We note this is appropriate since GRBs are generally found
to be close to the UV-brightest regions of their hosts \citep{fruchter06}, and the sizes of
high-$z$ galaxies are found to be intrinsically small \citep[e.g. 0.7$\pm$0.3\,kpc was found
for $z\approx7$--8 galaxies in the HUDF by ][]{oesch2010}. 
The prior knowledge of the exact locations of the GRBs allows us to 
adopt a \twosig\ excess flux as a reasonable detection threshold, which is unlike blind searches where
a much higher level of significance is required to both ensure confidence that the source is real
and also to reliably constrain the photometric redshift.
For comparison
to galaxy samples, and GRB hosts at other redshifts, it is relevant to consider the appropriate aperture 
corrections\footnote{i.e. the limits in this table are given by $M_{\rm lim} = 23.9-2.5 \log{(\mathrm{flux} + 2\sigma)} + \mathrm{ap}_{corr}$. The 
aperture corrections are calculated from the encircled energy (e.g. http://www.stsci.edu/hst/wfc3/documents/handbooks/currentIHB/ c07\_ir07.html)
and have values for WFC3/IR of $\rm ap_{105} = 0.31,\rm ap_{110} = 0.33, ap_{125} = 0.39, ap_{160} = 0.52$, for ACS/WFC $\rm ap_{850} = 0.27$ \citep{sirianni05}, and
for NICMOS/NIC3 $\rm ap_{160} = 0.60$. http://www.stsci.edu/hst/nicmos/documents/handbooks/ current\_NEW/c04\_imaging.6.5.html }, and these are included in the limits
shown in Table~\ref{photdata}.
Our errors are determined in each case from the variance within a set of  apertures of equal size to our source aperture, placed at random blank sky locations 
in the field surrounding the source. 

The results of the photometric analysis, and inferred limits on host properties,
are given in Table~\ref{photdata}.
Note that we report the measured sky-subtracted flux densities
even when they
are not significant detections, or indeed are formally negative.
This is to allow comparison with models, as is done in Section 3.

\subsection{GRB\,060522}

\label{sec060522}

Spectroscopy of GRB\,060522 was performed by Keck roughly 12.5\,hours after the burst, and revealed a strong break $\sim 7425$\,\AA~ in the spectrum, interpreted as
Ly$\alpha$ at $z=5.11 \pm 0.01$ \citep{cenko06}. A search for the host galaxy  with the {\em Spitzer Space Telescope}
found no detection
to flux densities of 0.2 $\mu$Jy at 3.6 $\mu$m and 2.4 $\mu$Jy at  5.8 $\mu$m, implying that its rest-frame optical luminosity lies substantially below \lstar\ 
\citep{chary07}.

We obtained observations of GRB\,060522 with WFC3/IR using the F110W filter. 
To ascertain the precise location of the burst on our {\em HST} image we performed relative astrometry between our WFC3 observations
and an image obtained at the Telescopio Nazionale Galileo (TNG) on 22 May 2006 \citep{davanzo06}. 
At the location of the afterglow we measure a flux density of 
$7 \pm 4$\,nJy, corresponding to a \twosig\ (aperture corrected) magnitude limit of F110W(AB)~$>28.13$. 
Although formally this is a non-detection at \twosig\, it is possible that some host flux is
contributing within the aperture. 
Indeed visually it appears there is a somewhat more significant
excess of flux which is offset slightly south-west from the GRB location (by about 0\farcs3, or $\sim2$\,kpc
at $z=5.11$), and
if we increase the aperture size to 0\farcs6 diameter then the flux density becomes
$12 \pm 5$\,nJy. 
This would represent a marginal detection, and corresponds to an apparent
magnitude of F110W(AB)$\approx$28.5.
While this is a plausible magnitude for a host (corresponding to
$M_{1800}({\rm AB})\approx-18$), 
the  probability of
a chance alignment with an unrelated object at such faint
magnitude levels is also non-negligible.
Specifically, considering the F110W number galaxy number counts
from \citet{thompson99} we estimate $\approx5$\% of random locations
will have a galaxy of magnitude F110W(AB)$\lesssim28.5$ within 0\farcs5.
Thus in a sample of six, as we have, there is a $\sim25$\% chance
of an unrelated galaxy being close enough to be possibly mistaken
for a host in at least one case.  For the sake of consistency with the other bursts
we will continue to use the 0\farcs4 aperture limit in the remainder of the paper.

\subsection{GRB\,060927}
GRB\,060927 was detected by {\em Swift}, and its optical afterglow was initially found in
rapid, but unfiltered Rapid Optical Transient Source Experiment (ROTSE) observations. An extensive 
follow-up campaign
\citep[reported in][]{ruizvelasco07}
revealed the source to be an $R$-band drop-out, well detected in the $i$-band. A subsequent spectrum
obtained at the Very Large Telescope (VLT) showed a faint continuum redward of $\sim 8000$\,\AA, as well as a weak
Si\,{\sc II} absorption feature, consistent with a redshift of $z=5.47$.

Our {\em HST}/NICMOS observations of the field of GRB\,060927 were 
taken using the F160W filter, and
were reduced as described in \citet{fynbo05}. 
GRB\,060927 was additionally re-observed with WFC3/IR using the F110W filter.
Astrometry was performed via observations made  with the VLT on 30 September 2006. 
To establish the position on the NICMOS images (which have a narrower field of view than WFC3) 
we opted to perform relative astrometry directly to the WFC3 images. 
The resulting \rms\ astrometric accuracy is $<$0\farcs01, and so does 
not impact the overall error in the astrometric solution described above. 

At the location of the afterglow the photometry in each of the NICMOS and WFC3/IR observations yields limiting 
magnitudes of F160W(AB)~$>27.75$ and F110W(AB)~$>28.57$  respectively.

\subsection{GRB\,050904}
GRB\,050904 was the first GRB at $z>6$ to be located. Its optical/IR  afterglow was initially 
found by \citet{haislip06}, who derived a photometric redshift of $z=6.39$ based on 
the strong spectral break between the and $i$- and $z$-bands, coupled with
nIR observations showing a blue spectral slope redward of the break \citep[see also][]{Tagliaferri05}. The afterglow was intrinsically
extremely bright, amongst the brightest observed for any burst, making late time
spectroscopy feasible with Subaru. This provided a measurement of the absorption redshift,
hydrogen column density, and metallicity of the host \citep{kawai06,totani07}. 
It remains the most distant burst for which all of these diagnostics are available. 

A search for the host galaxy of GRB\,050904 was conducted by  \citet{berger07} using
both {\em HST} and {\em Spitzer}, which placed deep limits on any host emission.
The first {\em HST} epoch, at which time both ACS/F850LP and NICMOS/F160W images were obtained,  was carried out
only about three weeks post-burst. No significant flux was detected in F850LP,
and the faint detection in the F160W image was shown to be due to residual afterglow
contamination since it was absent in a 
later F160W epoch \citep{berger07}.

For our analysis, we re-reduced the ACS data for GRB\,050904 (since we are primarily 
interested in the rest-frame UV luminosity close to Ly$\alpha$,
this filter is the more relevant). 
Astrometric tying of the afterglow to field sources was done utilizing a $z$-band image obtained from Gemini-South on 
7 September 2005.
At the location of the afterglow in the F850LP frame we measure a flux density of 
$-5 \pm 17$\,nJy,
corresponding to a \twosig\ limit of F850LP(AB)~$> 27.50$ (or 27.06 at $3\sigma$). 
This is 
in good agreement with
the limits reported by \citet{berger07}. 
However, since the Ly$\alpha$ break lies within the filter bandpass
at this redshift,
we must account for flux lost due to IGM absorption (i.e., the effective filter
width is narrower for this host), and so we conclude a corrected 
magnitude limit of F850LP(AB)~$> 26.86$.

\subsection{GRB\,080913}
GRB\,080913 was identified as a high redshift candidate based on photometric observations 
with GROND showing the burst to be an $i$-band drop out
\citep{rossi08}. 
Deep, red spectroscopy from the VLT  showed a strong spectral break, interpreted as 
Ly$\alpha$ at $z\sim 6.7$ \citep{greiner08b}. 
A further detailed analysis of the spectrum by \citet{patel10} revealed a single 
absorption line of Si\,{\sc II} at $z=6.733$, and we adopt this as the redshift of GRB\,080913.

We obtained 
{\em HST} observations with WFC3/IR in the F160W filter.
To tie the  astrometry of our {\em HST} observations we utilized images obtained from FORS2 at the VLT on 13 September 2008 in the $z$-band, and images
taken with NIRI on Gemini-North on 14 September 2008. 
We used 
two images independently to confirm the precise location of GRB\,080913 on our {\em HST} images. For the FORS2 images we identified 9 compact
sources in common, while only 6 sources were usable from the Gemini observations.
Although the afterglow detections are of low S/N, the 
errors on their centroids are small in comparison to
the errors derived from the fit 
and hence we are able place the afterglow to a \rms\ 
accuracy of 0\farcs08. Again, no host galaxy is visible in our observations to a limit of F160W(AB)~$>27.92$.

\subsection{GRB\,090423}
GRB\,090423 was first identified as a candidate high-redshift object based on its afterglow being a $Y$-band dropout, which
implies  $z>7.5$ \citep{cucchiara09}. 
VLT spectroscopy with both ISAAC and SINFONI 
allowed the identification of the Lyman break, despite low S/N, 
establishing the redshift $z=8.23 \pm 0.07$ 
\citep{tfl09}. This value
is in excellent agreement with the $z=8.1^{+0.1}_{-0.3}$ determined from a spectrum obtained at the TNG \citep{sdc09}.

We obtained our first {\em HST} observations of GRB\,090423 on 24 January 2010. 
At this stage we acquired 20 orbits
in each of the F125W and F160W filters on WFC3. Unfortunately 15 of these 20 orbits were
substantially impacted by persistence from earlier observations of bright field sources. This affected the
sensitivity over wide regions of the detector, although does allow us to measure the flux at the location
of the GRB host (which was only mildly affected by persistence). Additional observations were obtained
in October 2010. All of the available observations in each filter were co-aligned and stacked via {\tt multidrizzle}. 
These observations are extremely deep, with
limiting magnitudes a factor 1.5$\times$ deeper than the WFC3 Early Release Observations of the 
{\em Chandra} Deep Field South/Great Observatories Origins Deep Survey (CDF-S/GOODS) fields, 
and only another factor of 1.5$\times$ shallower than the WFC3 observations of the HUDF \citep{Bouwens11a}

The location of the afterglow on our
{\em HST} images was achieved by tying the astrometry to 
VLT/HAWK-I observations that were obtained approximately 17 hours post burst. While the afterglow had
faded since its first discovery from Hawaii, these deep observations yielded similar signal-to-noise, with
the added advantages of a larger field of view, and a greater number of faint sources for comparison.

Within the {\it HST} image there is no obvious source at the afterglow location, the
measured \twosig\ limiting magnitudes are F125W(AB)~$>30.29$, F160W(AB)~$>28.36$.
Since both filters sample the rest-frame far-UV, we form a weighted average of the
two results to provide a combined flux density measure of $-0.15\pm1.7$\,nJy, which
is used to derive the limit on the SFR reported in Table~\ref{probs}.
Such a small host is consistent with the non-detection of molecular
gas from the GRB location \citep{stanway11}.

\subsection{GRB\,090429B}
GRB\,090429B is the only host within our sample that does not have a spectroscopically measured redshift. However, the photometric break
between the $J$- and $H$-bands, coupled with the blue spectral slope between $H$ and $K$ provides a best-fit photometric redshift of $z=9.4$ 
and a robust lower limit to the redshift of $z\gtrsim6.5$
\citep{Cucchiara11}. 
As we commented above, the smooth power-law spectra of GRB afterglows makes photometric redshift
estimates generally more reliable than they are for galaxies, since the range of intrinsic spectral variation is much less \citep[e.g.,][]{kruehler11}.
We obtained observations in F606W (ACS), F105W, and F160W;
the non-detection of a host galaxy in these images, including the blue filters, provides additional support for the
high-$z$ origin for this burst, 
since the hosts of GRBs at $z<3$ have so-far always been detected
in {\em HST} optical imaging.
Here we consider primarily the F160W observation, since this is the only filter redward
of Ly$\alpha$ at the best-fit redshift  $z=9.4$.  
However, for completeness we also report results for  the $z\approx6.5$ lower limit, which allows us
to use both the F105W and F160W data.

We ascertained the location of the burst on the {\em HST} images via relative astrometry between our first epoch $K$-band observations and
those obtained with {\em HST}. 
The \twosig\ 
limiting magnitude at this location is F160W(AB)$>27.78$. This is shallower than for the majority of the bursts in our sample, since only
a two orbit exposure was obtained.  However, despite this the image still 
probes to faint limits comparable to the likely characteristic galaxy luminosity, \lstar, at $z=9.4$.

\section{Analysis and Discussion}

\subsection{Limits on host properties}
\subsubsection{Star formation rates}
\label{sfr}
At the redshifts of the bursts in question our nIR observations probe rest-frame wavelengths roughly in the range 1300-2000\,\AA. 
Limits on the UV luminosity at these wavelengths provide direct constraints on the host star-formation rates. 
A potentially important consideration in the UV is the effect of extinction by dust, which
could lead to a significant underestimate of the true SFR if uncorrected.
However, observational constraints from GRB afterglows at high redshifts \citep{greiner08b,tfl09,Zafar11} 
suggest that extinction corrections are likely to be small.
Similarly  the blue colours
of the $z\sim7$ candidate galaxies identified in the deep {\em HST} fields (e.g., \citet{bouwens10b}, \citet{finkelstein11}, but see also \citet{mclure11})
also argue for little dust in most early star-forming galaxies.
We therefore assume dust extinction can be neglected.

Following \citet{madau98} we estimate the 
star formation rates based on the UV luminosity at $\sim$1500\,\AA\ \citep[see also discussion in][]{bunker10}:

\begin{equation}
SFR = {L_{1500,UV} \over 8 \times 10^{27}\,\mathrm{ergs~s^{-1} Hz^{-1}} }\,M_{\odot} \mathrm{yr^{-1}}
\label{SFR}
\end{equation}

\begin{figure*}[ht]
\begin{center}
\resizebox{12truecm}{!}{\includegraphics{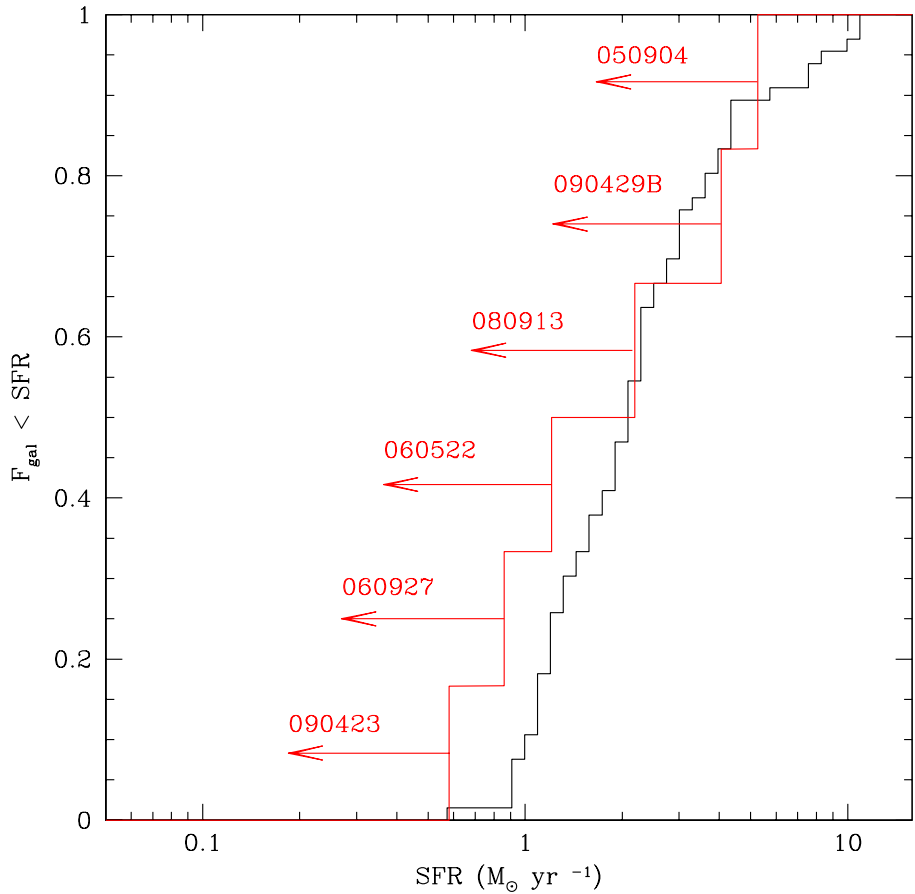}}
\end{center}
\caption{A cumulative histogram of the star formation rates (based 
on the 1500\,\AA\ magnitude) of a sample of $z \sim 7$ galaxies \citep[from][where
we have assumed each galaxy lies at $z=7.1$]{Bouwens11a} compared with the inferred 
\twosig\ limits on the star formation rates from our observations of GRB hosts. As can be 
seen, the limits attained in several individual cases correspond 
to galaxies in the faintest third of the high-$z$ galaxy distribution (i.e. limits
in the case of all our WFC3 observed hosts). This confirms that much high$-z$ star formation is taking place in relatively faint galaxies, too faint
to be found in most flux limited surveys.}
\label{fig2} 
\end{figure*}

As discussed in Section~\ref{sec060522}, 
there is a marginal detection of what may be the host of  GRB\,060522 at $z=5.11$, slightly offset from the burst position,
but otherwise none of the hosts are significantly detected.
The inferred \twosig\ limits on their star formation rates 
are given in Table~\ref{probs},
and in Figure~\ref{fig2} are plotted as a cumulative histogram of upper limits, compared to the SFRs
for a sample of $z\sim7$ HUDF galaxies from \citet{Bouwens11a}. The limits span the range $0.4-4$\,M$_{\odot}$ yr$^{-1}$,
and indicate that the total star formation rates in these galaxies are modest.
For comparison, the median SFR for GRB hosts at $z\lesssim1$ are found to be around 1--2\,M$_{\odot}$ yr$^{-1}$ \citep{svensson10}.

\subsubsection{Stacked limits on host galaxy emission}
The observations of six high-$z$ fields provide the opportunity to stack the resulting data in an attempt to provide either a combined detection, or
composite limit on the observed host luminosity. However, this is complicated by the use of different instrument and filter combinations to 
obtain the imaging. Therefore, we do not attempt to include data taken with different instruments,
and consider
a stack of {\em only} the WFC3 observations.  
Thus, we omit the observations of GRB\,050904, and the NICMOS observations 
of GRB\,060927. 
To perform the stack we first re-scaled the individual images (taken in different filters)
such that the units of each image can be considered equivalent. 
We then computed the necessary offsets to overlay
the burst positions, and re-drizzled onto a single output frame with these offsets applied. 
Simple aperture photometry at this
location was then performed, with our errors based on the scatter in background apertures as before. 
The mean luminosity distance is 68500\,Mpc, 
corresponding to a mean redshift of $z=6.82$.
There is no significant excess flux at the afterglow location,
and formally the measured stacked flux density is 
$1 \pm 3$\,nJy. 
Hence, the mean flux density of each host galaxy is constrained to be
$0.2 \pm 0.6$\,nJy, 
corresponding to a \twosig\ limit of $m_{\rm AB} > 30.7$, equivalent to an absolute magnitude $M_{\lambda/(1+z)} < -16.2$ and 
SFR $<$0.17 M$_{\odot}$ yr$^{-1}$, at this mean redshift.

\subsection{Constraints on high redshift galaxy luminosity functions }
\label{glf}

Although our small sample of high-$z$ GRBs does not yet allow us to place strong constraints
on the galaxy LF, we can test whether the limits on the  host magnitudes are consistent with
them having been drawn from the LFs suggested by other studies of galaxy populations with 
redshifts between 5 and 10.

The recent re-observations of the Hubble Ultra-Deep Field with WFC3/IR have revealed a population of $z$-band
and $Y$-band dropouts, with colours consistent with galaxies at $z>7$ and $z>8$ respectively \citep{bunker10,mclure10,Bouwens:2010ly}. 
A single
candidate $z \sim 10$ galaxy ($J$-band drop out) has also been identified \citep{bouwens11}. 
From these samples, assuming they are substantially complete and uncontaminated, 
it is possible to make some statements about the form and 
evolution of the galaxy LF from $5 < z < 10$. 
In particular, these authors fit their data in bins of redshift with  the LFs  described by a Schechter function \citep{Schechter76}:

\begin{equation}
\phi(x)dx=\phi^*x^{-\alpha}e^{-x}dx
\label{schechter}
\end{equation}

\noindent
where $x=L/L^*$, with \lstar\ being the characteristic luminosity of the ``knee'' of the LF.
Here $\alpha$ is the power-law slope towards faint luminosities and $\phi^*$ is a normalisation factor.
Of course, there are no strong observational reasons to expect the LF to have this form at high-$z$,
but it is supported by theoretical work \citep{trenti10}.

\begin{figure*}[ht]
\begin{center}
\resizebox{13.5truecm}{!}{\includegraphics{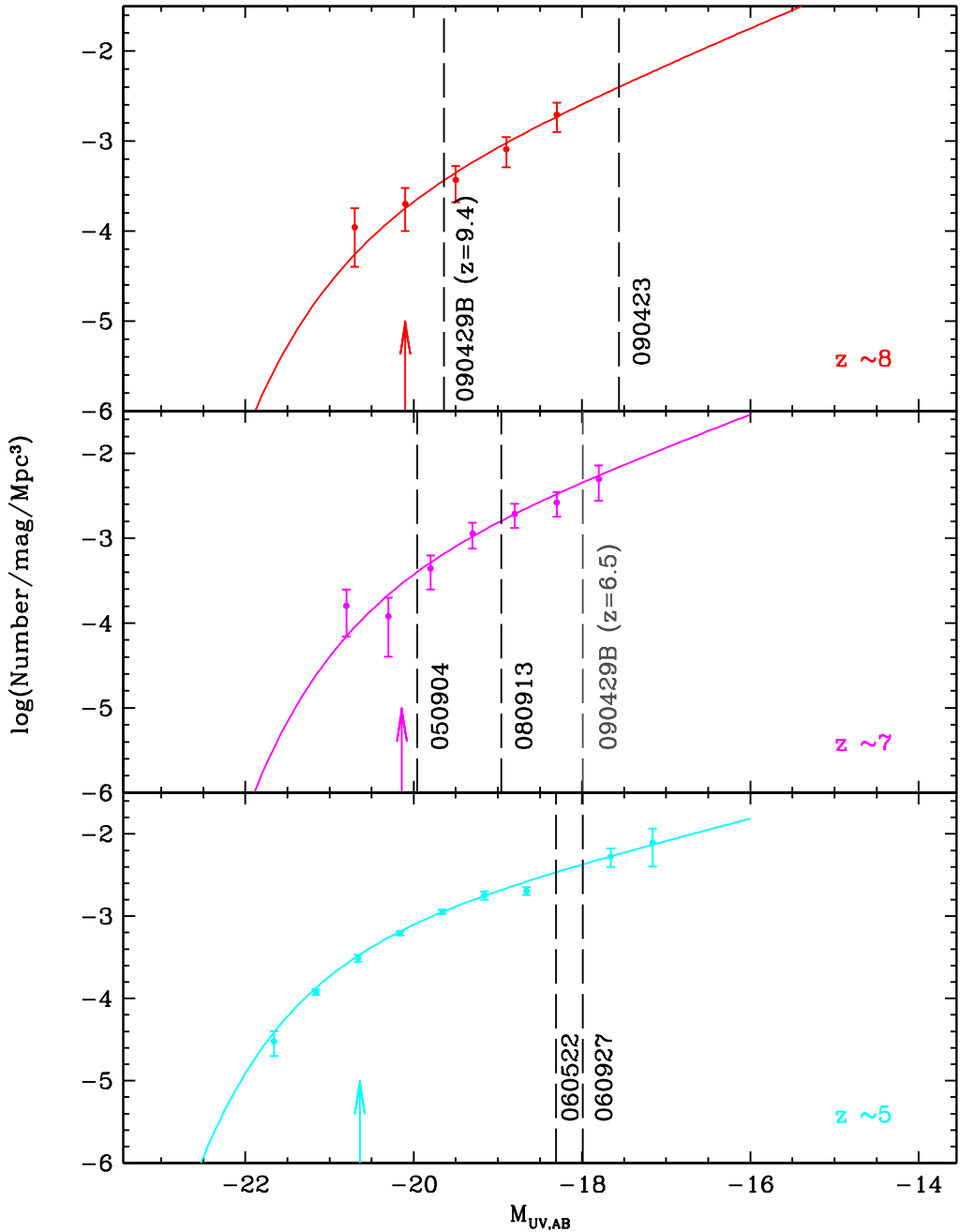}}
\end{center}
\caption{Our $2\sigma$ 
upper limits on the magnitudes
our individual host galaxies (vertical dashed lines), compared to
the luminosity functions of galaxies at $z=5,7,8$ (solid curves), as determined by
\citet{Bouwens11a}. 
In particular for GRB\,090423, the power of deep {\em HST} observations to probe sources at known
locations and redshifts to well below \mstar\ (indicated by upward arrows) can be readily appreciated.  
Note that GRB\,090429B is plotted both in the top, highest redshift, panel appropriate for its best-fit
photometric redshift, and also (lighter shade) in the middle, intermediate redshift, panel appropriate for
the lower-limit of photo-$z\approx6.5$ (see text for more details).
}
\label{fig3} 
\end{figure*}

In Figure~\ref{fig3} we show the preferred analytical LF fits of  \citet{Bouwens11a} and the data on which they are based, for samples
at $z\sim5$, 7 and 8.
Their conclusion is that the characteristic knee in the luminosity function, \lstar, becomes slowly
fainter with increasing redshift from 4--8, and at the same time the faint end slope becomes steeper, reaching
a value of $\alpha\sim2$ by $z=8$.
This is in addition to the overall normalisation, $\phi^*$, lowering  
\citep[but see also][who find a fading $L^*$ and lower $\phi^*$ but less evidence for a steepening slope]{mclure10}.
This indicates that an increasing proportion of star formation is occurring in fainter galaxies, and indeed formally 
the integrated luminosity represented by a Schechter function with $\alpha>2$ diverges without some lower
cut-off luminosity. 
However there are important caveats which pertain to these analyses: firstly, the LF parameters are based on entire samples, and
any incompleteness or contamination (particularly difficult to rule out at the faint end) will introduce biases;
secondly,  even the HUDF is limited to finding galaxies in the top few magnitudes of
the LF, so that  the
conclusions about total star-formation rate (and hence the production rate of ionizing photons) 
are sensitive to the untested assumption that a Schechter
function is the appropriate form, and to the large uncertainties on the measurement of the faint-end slope.

We note 
that at high redshift the UV luminosity function 
has limitations as a way of representing the whole population of galaxies,
since they are typically only visible when in the starbursting phase, and likely remain in this state for only a 
short duration based on the typical ages, and apparent availability of molecular gas for star formation within them. 
However, particularly for understanding the contribution of galaxies to the reionization of the Universe,
the measured UV luminosities, which  are more representative of star formation occurring with the last $\sim 15-100$\,Myrs,
are the relevant quantities. 
Furthermore, since the GRB itself is a sign of ongoing massive
star formation within the host, one would expect the likelihood that a GRB
occurs to scale with the UV luminosity.
Indeed, we would expect the region immediately underneath the burst to be 
extremely UV-bright, as is observed in the more local GRB population \citep{fruchter06}. 
Hence we can meaningfully compare 
our limits on host emission with the observed galaxy LFs at $z \sim 7$.

The   \twosig\ luminosity limits for the GRB hosts are also shown graphically in relation 
to the high-$z$  galaxy  LFs  
 in Figure~\ref{fig3}, 
while their tabulated UV absolute magnitude limits are given in Table~\ref{photdata}. As can be seen, all the hosts are
apparently 
below \mstar\ (the AB magnitude corresponding to \lstar)
at their respective redshifts, and in the case of our observations of GRB\,090423 have
the ability to probe to fainter limits than has so far been possible with the Ultra-Deep Field observations.
This is because, despite the shallower depth of our images,
we have prior knowledge that
an object exists at this location, and so can accept a  lower formal significance level for detection
(since the chances of a random noise fluctuation are lower due to the much smaller area under consideration). 
Furthermore,
since we do not need a blue ``veto" filter we make more efficient use of the available exposure time; 
in contrast to some Lyman break searches which are limited by the  depth of their short-wavelength (blueward of Ly$\alpha$)
imaging.

These observations of GRB host fields allow us to test the validity of the galaxy
LFs that have been derived from deep field observations. 
We assume that there is no dependence of GRB rate or luminosity on
environmental parameters such as metallicity, which, as discussed in Section~\ref{intro},
is certainly plausible at high
redshifts when the bulk of star-forming galaxies were generally not highly enriched.
We also assume that any dust extinction is minor, as argued in Section~\ref{sfr}.
Then the probability that a galaxy produces a GRB in some unit time (the probability of more
than one GRB in realistic observing times being negligible)
is approximately proportional to its rest-frame UV flux, 
since massive stars are responsible
for producing both: i.e., $P_{\rm GRB} \propto{\rm SFR}\propto{L_{\rm UV}}$.
Thus, because we would expect GRBs to be drawn randomly from the total 
stellar UV luminosity, 
this means that
the host galaxies should be drawn from the luminosity-weighted
galaxy LF.

In practice, 
the best fit values for the faint end slope $\alpha$ and
characteristic luminosity \lstar\ at each redshift considered were determined from the fitting
formulae given by \citet{bouwens1105.2038}, and are summarised in Table~\ref{probs}.
We can then calculate a luminosity-weighted luminosity function, which (suitably normalised)
we take to be
equivalent to a probability density function (PDF) for the intrinsic host luminosity:

\begin{equation}
y(L) =  {L \phi_z(L)  \over \int_{L_{\mathrm{min}}}^{\infty} L^{\prime} \phi_z(L^{\prime})dL^{\prime}} 
\label{PDF}
\end{equation}

\noindent
where $\phi_z(L)$ is the luminosity function at redshift $z$.
We emphasize again that, since we are going to compare to the fluxes measured in apertures at the exact
GRB location, we are also assuming there is no significant offset between the GRB and
its parent galaxy.  

Hence, the probability of observing a GRB in a galaxy of luminosity $L < L_{\rm host}$ can be 
obtained via the cumulative probability density function (CDF):

\begin{equation}
Y(L_{\rm host})=P(L<L_{\rm host}) = {\int_{L_{\mathrm{min}}}^{L_{\mathrm{host}}} L^{\prime} \phi_z(L^{\prime})dL^{\prime}  \over \int_{L_{\mathrm{min}}}^{\infty} L^{\prime} \phi_z(L^{\prime})dL^{\prime}} 
\label{CDF}
\end{equation}

Setting a lower limit, $L_{\mathrm{min}}$, for the integral is physically motivated
since small dark halos ($\lesssim10^8M_\odot$) at $z\sim10$ are expected to retain little gas
and form few stars \citep{read06}.
Given the apparent steepness of the faint end slope of the LF at $z \sim 8$ it
is also important that the value of $L_{\mathrm{min}}$ be sensibly chosen. 
It is not feasible to directly
measure this lower limit at high redshift 
since it is well below the detection threshold
for deep imaging, and will remain so even in the era of the {\it James Webb Space Telescope (JWST)} or ground-based 
Extremely Large Telescopes (ELTs). However, it is possible 
to provide estimates via simulations \citep{read06}, or from
the star formation histories of the lowest mass galaxies in the local universe \citep[e.g.][]{weisz11}. 
This latter approach suggests that the lowest mass galaxies attain total stellar masses
of $\sim 10^6$ M$_{\odot}$, over a Hubble time, implying mean star formation rates of 
$\sim 10^{-5} - 10^{-4}$ M$_{\odot}$ yr$^{-1}$. However, star formation 
is likely to be episodic,
and characterized by periods when the SFR is markedly
higher than this average, and other times when the star formation is inactive, and the galaxy effectively 
invisible in the UV. Indeed, the study of \citet{weisz11} suggests that
the SFR of local
dwarfs 
 was somewhat higher in the early Universe than their average SFR over the
Hubble time. On the basis of this we adopt a cut-off value of 
$L_{\mathrm{min}} = 4 \times 10^{23}$\,erg\,s$^{-1}$Hz$^{-1}$
which is equivalent to $M_{\rm AB}=-10$, 
and is similar to that considered by other recent studies \citep{bouwens1105.2038,kuhlen12}.
This corresponds to a star formation rate of 
$\approx 5\times10^{-4}$ M$_{\odot}$ yr$^{-1}$.
We note, that when $\alpha < 2$
(i.e., for redshifts less than $z\sim7$) the precise
choice of $L_{\mathrm{min}}$ has little impact on the results. 

\subsubsection{Analysis I }

We now consider the question: what is the probability that each host individually is 
fainter than the $2\sigma$ upper bound we have inferred for its luminosity?
To this end, we calculate $P_{\mathrm{2\sigma}}$ for each host by setting $L_{\rm host}$ in Equation~\ref{CDF} equal to this upper 
bound.  

\begin{figure*}[ht]
\centerline{\includegraphics[width=8cm,angle=270]{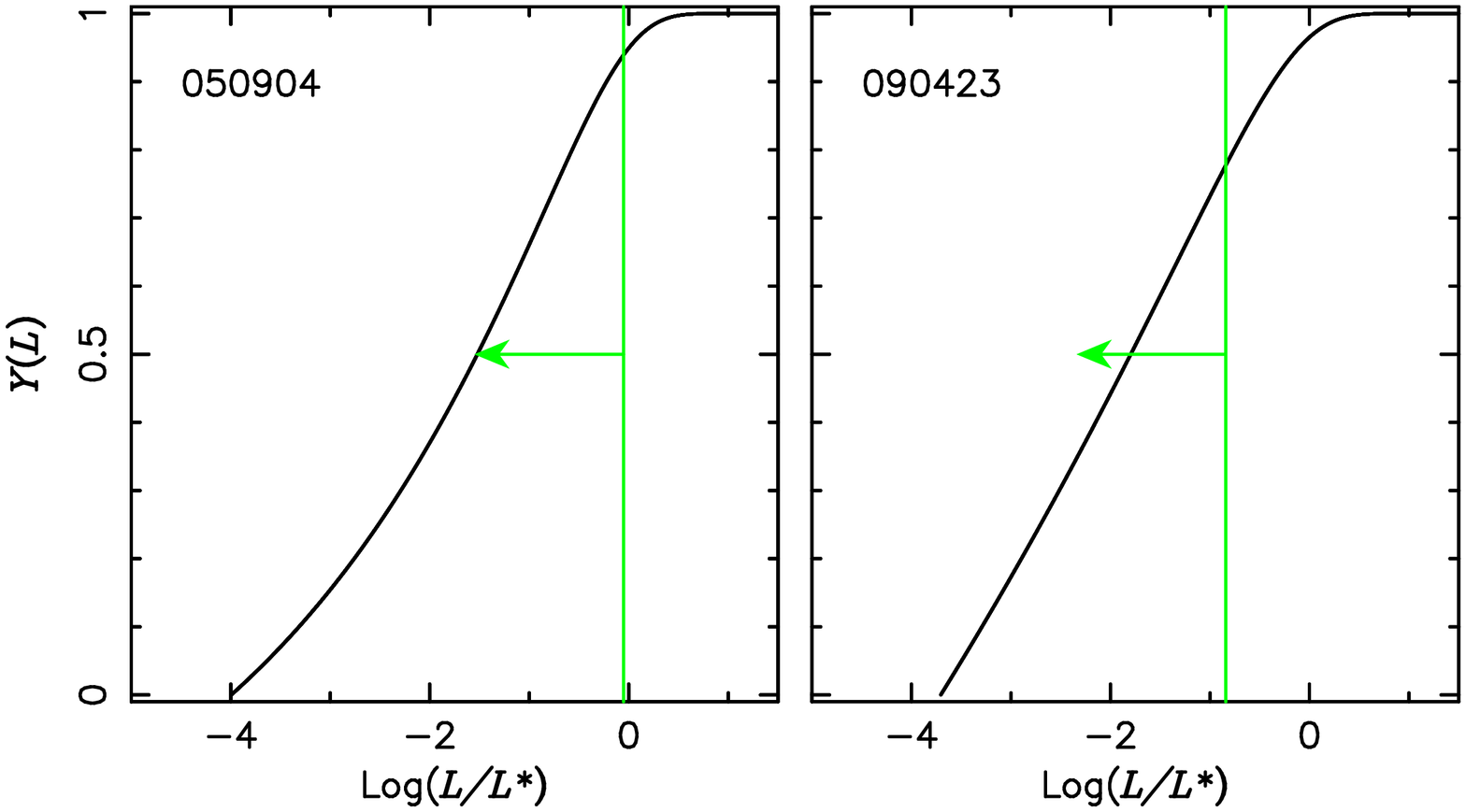}}
\caption{
These panels show
 the cumulative probability distributions for the luminosities of
 two of our GRB hosts,
based on assuming GRB likelihood is proportional to UV luminosity,
as described in Section~\ref{glf}.
Galaxy LF parameters are taken from \citet{bouwens1105.2038}.
The  vertical lines 
correspond to the \twosig\ detection limits for our {\em HST} images.
As noted, in no case do we detect a host at this level of confidence.
}
\label{fig4} 
\end{figure*}

The results are summarised in Table~\ref{probs} and
illustrated in Figure~\ref{fig4}, which shows the CDFs for luminosity (expressed
relative to \lstar\ at the redshift in question) 
of two bursts from our sample,  
together with
the \twosig\ detection limits for the corresponding {\em HST} frames.
We see that, given the above
assumptions, it is not surprising to find individual hosts to be undetected since 
the majority of the likelihood lies below these \twosig\ bounds.
However, the joint probability that none of the hosts is detected is only
$P{\rm(none)} = \prod P_{\rm 2\sigma} = 0.17$.  While not a highly significant result, this does suggest
that a non-evolving LF, in which more star formation was taking place in galaxies with luminosities around \lstar,
would not sit comfortably with the apparent faintness of the GRB hosts.

\subsubsection{Analysis II}

A weakness of the above approach is that it does not make full use of the joint probability of the formally measured fluxes
at the positions of the whole sample of GRB hosts.
As an alternative, we perform the following analysis, again with the aim of testing
the evolving galaxy LFs proposed by 
\citet{bouwens1105.2038}.

As before, we construct a CDF for host luminosity, but now
turn this  into an equivalent CDF for observed flux density $F$ using the cosmological
luminosity distance for the given redshift, having accounted for the aperture corrections.
The next step is to convolve this with the observational errors appropriate
for the given GRB field observation, in order to obtain a CDF for the {\em observed} host galaxy flux density:

\begin{equation}
Y(F) = {  \int_{F_{\mathrm{min}}}^{F_{\rm obs}} G * (F^{\prime} \phi_z(F^{\prime}))dF^{\prime}  \over \int_{F_{\mathrm{min}}}^{\infty} G * (F^{\prime} \phi_z(F^{\prime}))dF^{\prime} }
\label{PDF2}
\end{equation}

\noindent
where $G$ is a gaussian with a width $\sigma$ dictated by the sky noise in each image, measured
from numerous sky apertures of equal radius to the source aperture. 
$F_{\mathrm{obs}}$ is the formal flux measurement at the location of the GRB, and
the minimum flux density, $F_{\mathrm{min}}$, is appropriately scaled from the minimum luminosity, discussed above.

\begin{figure*}[ht]
\centerline{\includegraphics[width=8cm,angle=270]{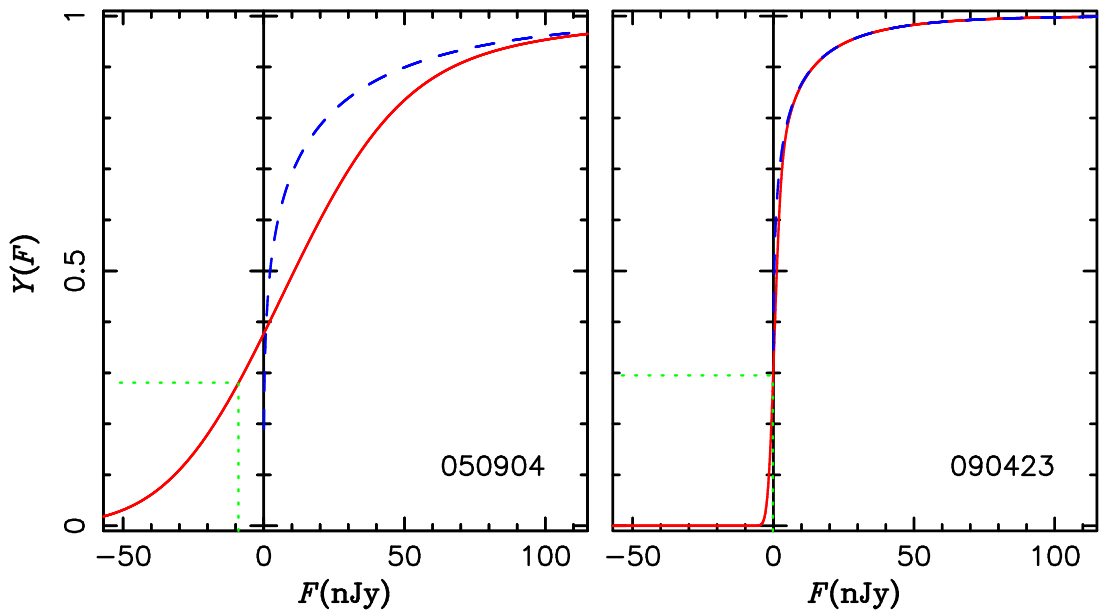}}
\caption{
Two examples of the
cumulative probability density 
functions for the  GRB host {\em true} flux densities (dashed curves; blue on-line), based on
the luminosity-weighted galaxy LFs, as described in Section~\ref{glf}.
These are then convolved with the observed measurement errors for each field
to produce the solid (red on-line) curves which are the predicted CDFs for the {\em measured}
flux densities.  The dotted (green on-line) lines show the formal measured flux density at the location of the GRB in each case.  
Reading these across on the CDF axes we would expect to find
locations drawn randomly from a uniform distribution between 0 and 1.
}
\label{fig5} 
\end{figure*}

For illustration, the CDFs for two of our bursts
are shown in Figure~\ref{fig5} (red curves), along with the 
CDFs for true host flux for comparison (blue curves).
The green lines indicate the formal measured flux density at each GRB position.
Note, that for GRB\,090423 we take a weighted average of the results for both
filters, since they straddle 1500\,\AA,
while in the analysis of GRB\,060927 we use the F110W flux density, being the closest match to 1500\,\AA.
If the assumptions we have made are correct, then we would expect these
measured flux densities to be drawn randomly and uniformly from 0 to 1 on the 
cumulative probability axes.  
We can quantify this by calculating the average value for $\langle Y(F_{\rm obs})\rangle = 0.46$,
consistent with an expected value of $0.5\pm0.12$ from the Central Limit Theorem.

\begin{figure*}[ht]
\centerline{\includegraphics[width=8cm,angle=270]{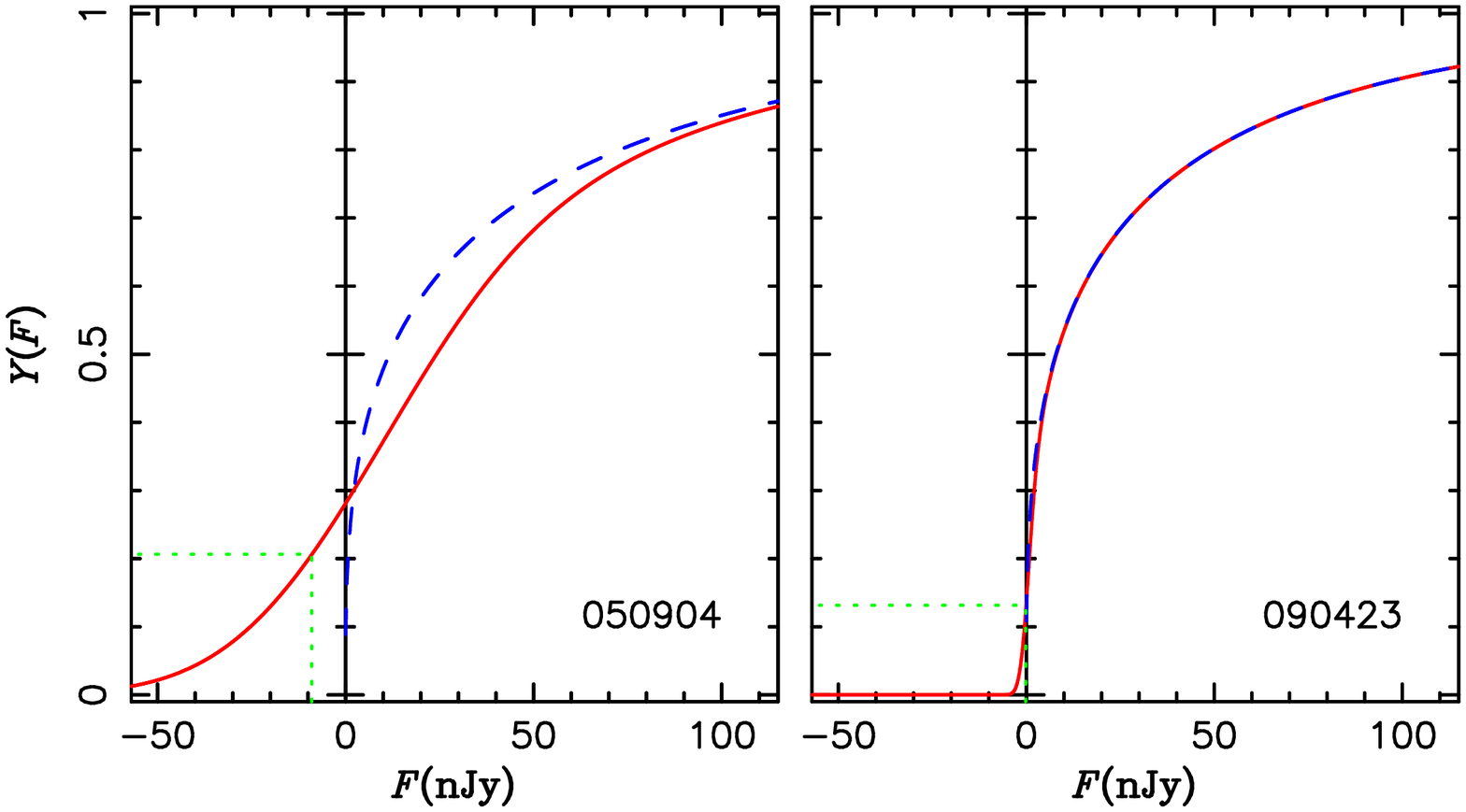}}
\caption{
As with Figure~\ref{fig5} but here we have fixed the shape of the input galaxy LFs
to $\alpha=1.73$, $M^*=-20.97$ which is representative of the population at $z\approx3$. 
The shallower slope and brighter characteristic luminosity of this LF leads to 
our faint flux limits corresponding to
systematically
lower values of the CDF, $Y(F)$.
}
\label{fig6} 
\end{figure*}

However, we can also repeat the analysis, but this time
fix the LF parameters for all the hosts to be that found at lower redshift.
Specifically, we first choose $\alpha=1.73$ and $M^*=-20.97$, 
as measured at $z\approx3$ by \citet{reddy09}.
Thus we are testing here whether the hosts could be drawn from a luminosity function
whose shape (but not normalisation) does not evolve from $z\sim3$ to high redshift.
The results in this case 
for the same two bursts
are shown in Figure~\ref{fig6}.
Now all the measured flux densities at the GRB locations are close to or below the $Y=0.5$ level,
with a mean $\langle Y(F_{\rm obs})\rangle = 0.29$,
thus rejecting the model at $\approx96$\% confidence 
($\approx98$\% if we took $z=6.5$ for the redshift of GRB\,090429B and averaged the
F105W and F160W limits).
An alternative test would be to fix the parameters to those found
at $z=6$--7 by \citet{mclure10}, $\alpha=1.71$ and $M^*=-20.08$.
Again, this model is weakly rejected at the $\approx90$\% level
($\approx94$\%).

Our result is not a highly significant, but does support an evolving galaxy LF, with an
increasing proportion of star formation occurring in faint galaxies.
It also demonstrates
that a larger sample and/or deeper limits on host emission can begin to provide important tests
of the high-$z$ galaxy LF.

\begin{deluxetable*}{lllccccl}
\footnotesize
\tablecolumns{4}
\tablewidth{0pt}
\tablecaption{Log of {\em HST }observations of the host galaxies of GRBs at $z>5$}
\tablehead{\colhead{Date} & \colhead{UT Time} & \colhead{Filter} & \colhead{$\lambda_{\rm rest}$(\AA)} &
\colhead{Exp (s)}& \colhead{$F_{\rm obs}$ (nJy)}
&  \colhead{AB mag limit}  & \colhead{M$_{\lambda/(1+z)}$} }
\startdata
{\bf 060522} \\
17 Oct 2010 & 10:30 & F110W & 1888 & 8395 & $7 \pm 4$ & $>28.13$  & $>-18.35$  \\
\hline
{\bf 060927} \\
29 June 2007 & 11:30 & F160W & 2376 & 10240 & $7 \pm 5$ & $>27.75$  &$>-18.84$\\ 
25 Sept 2010 & 14:30 & F110W & 1782 & 13992  & $4 \pm 3$ & $>28.57$ &$>-18.02$ \\
\hline
{\bf 050904} \\
26 Sept 2005 & 21:03 & F850LP & 1279$^\dagger$ & 4216 &  $-9\pm30$$^\dagger$  & $>26.86$ & $>-19.95$ \\
\hline
{\bf 080913} \\
30 Nov 2009 & 16:10 & F160W & 1988 &7818 &$3 \pm 6$ & $>27.92$ & $>-19.00$ \\
\hline
{\bf 090423} \\
24 Jan 2010 & 11:34 & F160W & 1665 & 13029 &  \\
25 Jan 2010* &  14:44 & F160W & 1665 & 13029 & $4 \pm 3$ & $>28.36$  & $>-18.88$\\  
26 Jan 2010* &  13:06& F125W & 1353 & 13029 &  \\
27 Jan 2010* &  13:04 &  F125W & 1353 & 13029 &   \\
22 Oct 2010  &  18:23 & F125W & 1353 & 13029 & \\
27 Oct 2010 &   16:36 & F125W & 1353 & 13029 & $-2\pm 2$ & $>30.29$ & $>-16.95$ \\  
\hline 
{\bf 090429B}  \\
10 Jan 2010 & 21:54 & F160W & 1478 & 2412 &   &  &  \\
22 Feb 2010 & 19:22 & F160W & 1478 & 2412 & $7 \pm 5$  & $>27.78$ & $>-19.65$ \\
                        &            &               & (2049)&          &                     &                    & $(>-19.09)^\spadesuit$ \\
24 Feb 2010 & 03:19 & F105W & 1014  & 2412&                     &                    &  \\
28 Feb 2010 & 13:56 & F105W & 1014 & 2412 & $-1 \pm 5$  & $>28.49$ & $>-18.73$ \\
 &  &  & (1407) &  &   & & $(>-18.17)^\spadesuit$ 
\enddata
\tablecomments{~Log of {\em HST}  observations of the locations of our sample of bursts giving
the details of the observations and the pivot wavelength of the filter at the assumed redshift, $\lambda_{\rm rest}$.
$F_{\rm obs}$ represents the measured (sky-subtracted) residual flux density
in an aperture of diameter 0\farcs4, centred on the position of the
afterglow. 
The \twosig\  point-source
limits on apparent magnitude (after aperture correction) and corresponding absolute magnitude 
for each filter are also given. 
*Images affected by persistence. {$^\dagger$}Filter central wavelength and
flux density calibration have been corrected for the fact that the Ly$\alpha$ break 
for GRB\,050904 falls roughly in the middle of the filter response, and so 
in effect only the longer wavelength part of the filter passband is actually
sensitive to the flux from the host, and the quoted limit is for the flux density redward of Ly$\alpha$.
$^\spadesuit$Result if GRB\,090429B was actually at the lower limit 
for its photometric redshift of $z\approx6.5$.
}
\label{photdata}
\end{deluxetable*}

\subsection{Implications for reionization}

Whether UV radiation from star-forming galaxies is sufficient to bring about
and sustain the reionization of the IGM above $z\sim6$ is a long-standing question \citep[e.g.][]{loeb09}.
Recently a number of authors  have argued that an increasingly steep 
faint end to the galaxy LF, motivated by theoretical considerations,
may be able to achieve this, without resorting to extremely high Lyman continuum escape fractions 
\citep[e.g.][]{bouwens1105.2038,lorenzoni11,kuhlen12}.
However, the form, steepness and faint-end cut-off of the galaxy LF at $z>7$
are very poorly constrained by current data, since {\em HST} (and even in the longer-term
{\em JWST} and ground-based 20--40\,m class optical/nIR telescopes) can only directly probe the bright end of the LF.

If GRBs are sampling star formation in an unbiased way, as we argue they may be
at early times, then they provide an alternative window on the total star formation rate
which would ultimately circumvent the necessity to detect the individual galaxies
in which the star formation is occurring.
Our results, even from the small sample of high-redshift bursts currently available,
already support an evolving galaxy LF over a non-evolving one, and therefore
suggest that reionization may be brought about primarily from stars born in very
faint proto-galaxies.

\subsection{Possible biases}
It is worth considering further possible physical effects which may be biassing our
conclusions.  If a significant amount of star formation at high-$z$ is actually dust enshrouded,
which we believe is unlikely, then it would impact on the observability of GRBs and their
hosts, as well as the LBG samples.  One would generally expect that the GRBs for
which afterglows are detected and redshifts estimated would typically be in the 
lower dust systems, if there is indeed a wide range, and so in that sense, their hosts  could
still be compared directly to the (also preferentially dust free) LBG samples.
From the point of view of reionization, of course it is the low-dust star formation that
is more likely to have a high escape fraction of ionizing radiation, so our conclusions
are likely to be valid in that respect.

As discussed in section 1, we should also be concerned about possible GRB metallicity sensitivity.  
Both theoretical
considerations and observational evidence (the blue colours of the LBG samples)
argue that few star-forming galaxies at $z>6$ will have high metallicities (e.g. supersolar)
and therefore the low rate of GRBs in such systems seen at low redshift is not
likely to be an important factor at high-$z$.  Of course, it could be that a contrary effect becomes
important at some point,  for instance if very low metallicity
populations produce fewer and/or fainter GRBs.  However, if that were the case
then we would expect to lose, if anything, the lower mass halos, where any metals
produced are most easily lost. Hence, such an effect would seem unlikely to result in
finding an unusually faint population of hosts, and so again our basic conclusion would
only be strengthened.

\section{Conclusions}

We have presented {\em HST} observations of a sample of six GRB host galaxies beyond $z \sim 5$. 
One host, that of GRB\,060522 at $z=5.11$ may be marginally detected, but the others are undetected
to deep limits, typically $H_{\rm AB}\sim28$ at \twosig.
If GRBs are good tracers of the locations of star formation at high-$z$
our results confirm that much, and probably 
the majority of, star formation then
was taking place in small galaxies that are too faint to be detected even in the various {\em HST} deep field surveys. 
While the sample is small,
the joint probability that none of the hosts are detected is consistent
with a galaxy LF which is  rapidly evolving to higher redshifts, and marginally inconsistent with
an LF whose shape does not evolve.

Our analysis does rely on two assumptions: that environmental conditions, such
as abundance variations, do not produce appreciable variations in the SFR to
GRB-rate ratio from galaxy to galaxy, and that dust content is generally negligible.  Both of these are consistent
with our current understanding of early galaxies, and of GRBs, but doubtless
work is required to further clarify these issues. In terms of the effect of
dust, we note that from the point of view of reionization, the un-corrected luminosity
function is really what we are interested in, since it is largely the unobscured star
formation which will contribute to the inter-galactic UV radiation field. However,
if there were significant dust in some GRB hosts, it would weaken the connection
between GRB-likelihood and apparent UV luminosity which we have assumed.

This work demonstrates the potential power of GRB-selected galaxy samples  to 
quantify the amount of star formation occurring in faint galaxies 
at early times,
which is essential for understanding
the budget of UV photons and their role in reionizing the Universe.
In the future, deep imaging of larger samples of GRBs at high-$z$, combined with 
better understanding of any environmental dependencies of GRB production,
could provide much more stringent constraints on
the faint-end of the galaxy luminosity function.
Specifically, such a sample could in principle be used to fit for all the LF shape
parameters, including the minimum cut-off luminosity, or, more empirically,
simply determine the relative SFR fraction in bright (detected) galaxies
compared to that in faint (undetected) galaxies, without any prior assumption about
the form of the LF.

Just as this paper was submitted, two other papers on the same topic
appeared as preprints.  \citet{basa12} have reported deep VLT observations
of three high-$z$ fields, of GRBs\,060522, 060927 and 080913.  The point-source magnitudes
reached are rather shallower than the {\em HST} limits, and 
their non-detections therefore consistent with our results.
\citet{trenti12} instead performed a theoretical analysis, aimed at predicting
the fraction of GRB hosts expected to be detected in deep {\em HST} imaging
at different redshifts, under various model assumptions.
For example, they predict that 50\% of GRB hosts at $z\sim5$ should be
detected in a survey reaching $M_{\rm AB}\sim-18$, consistent with our possible
detection of the GRB\,060522 host.  To the same rest-frame limit, they predict
fewer than 20\% of hosts should be detected at $z\sim8$, again, in agreement with
our findings.

Traditional blank-field imaging surveys,
while potentially finding larger samples of candidate galaxies, are
limited to detecting just the relatively brightest examples at the most extreme redshifts.
This may be true
even in the era of the {\em JWST},
especially given the apparent dearth of $z\sim10$ candidates located so far in the HUDF \citep{oesch12}, 
which could be indicating an even more rapid evolution of the galaxy LF parameters.
The approach we, and the other studies mentioned above, have adopted
provides a crucial complementary insight into early galaxy evolution.

\acknowledgements{}
We acknowledge support from STFC. Based on observations made with the NASA/ESA Hubble Space Telescope, obtained from the data archive at the Space Telescope Institute. STScI is operated by the association of Universities for Research in Astronomy, Inc. under the NASA contract  NAS 5-26555. These observations
are associated with {\em HST} programs GO-10616 (PI: Berger), GO-10926 (PI: Tanvir), GO-11189 (PI: Tanvir) and GO-11734 (PI: Levan). 

JPUF acknowledges support form
the ERC-StG grant EGGS-278202. The Dark Cosmology Centre is funded by the DNRF.

We thank Justin Read for useful discussions.

\begin{deluxetable*}{llcccccccc}
\footnotesize
\tablecolumns{4}
\tablewidth{0pt}
\tablecaption{Assumed  luminosity function parameters at each redshift, and derived host properties and probabilities}
\tablehead{\colhead{Burst} & \colhead{$z$} & \colhead{\mstar$_{UV,AB}$} & \colhead{$\alpha$} & 
\colhead{$m^*$} & \colhead{$F^*$(nJy)} & \colhead{$L_{\rm host}$/\lstar}&    \colhead{SFR(M$_{\odot}$yr$^{-1}$)} &
\colhead{$P_{\rm 2\sigma}$} & \colhead{$Y(F_{\rm obs})$}}
\startdata
060522 & 5.11 & -20.59 & 1.80 & 26.22 & 118 & $<0.09$ & $<0.88$ & 0.55 & 0.47\\
060927 & 5.47 & -20.49 & 1.81 & 26.43 &   97 &$<0.08$ & $<0.65$ & 0.55 & 0.43 \\
050904 & 6.29 & -20.26 & 1.85 & 26.83 &   67 & $<0.59$ & $<4.1$ & 0.94 & 0.28\\
080913 & 6.73 & -20.14 & 1.88 & 27.33 &   43 &$<0.21$ & $<1.3$ & 0.82 & 0.44\\
090423 & 8.23 & -19.72 & 1.95 & 27.97 &   24 &  $<0.09$ & $<0.38$ & 0.78 & 0.30\\
090429B & 9.4 & -19.39 & 2.01 & 28.59 &   13 & $<0.78$ & $<2.4$ & 0.96 & 0.82\\
      & (6.5)$^\spadesuit$ & (-20.20)&(1.87)&(26.68)&(77)& $(<0.13)$ & $(<0.84)$ & (0.71) & (0.46)\\
\vspace{-1mm}	
\enddata
\tablecomments{
Columns 3--6 give the galaxy luminosity function parameters assumed at the redshifts of each GRB 
obtained from the fitting formulae of \citet{bouwens1105.2038},  both
in absolute terms and translated to apparent values (including correction to finite aperture).
Columns 7--8 list the \twosig\ limits on host luminosity and 
star-formation rate; in the case of GRB\,060927 based
on just the F110W image, and for GRB\,090423 based on the weighted average
of both filters, as described in the text.
Finally, columns 9--10 give the calculated probability that the host would be below this
\twosig\ limit, $P_{2\sigma}$
and the position in the 
cumulative probability density function of the observed flux density, $Y(F_{\rm obs})$,  based on an evolving galaxy  LF (Section~\ref{glf}).
The combined probability of the $2\sigma$ upper limits, the product of the figures in column 9,
is $\prod{P_{2\sigma}}=$0.17.  
$^\spadesuit$Result if GRB\,090429B was actually at the lower limit 
for its photometric redshift of $z\approx6.5$.
}
\label{probs}
\end{deluxetable*}

\bibliographystyle{apj}
\bibliography{bibi090429B}

\end{document}